# A trampoline effect occurring in the stages of planetary reseeding

**Ian von Hegner**

Aarhus University

**Abstract** Impactors have hit the Earth since its formation and have continued to be infrequent guests throughout the Earth's history. Although the early part of the Earth's history was marked by these violent events, life was present early, possibly existing already in the Hadean Eon. It is possible that life has been, and still is, transported between the worlds of the solar system, owing to impacts leading material away from the impact region. Beyond this lithopanspermia theory, in the so-called 'refugium hypothesis', ejected material has been suggested to also return to its home planet and 'reseed' life after the world has recovered after a global impactor, thus restarting evolution. In addition to such impactors, more frequent impacts from smaller non-sterilizing impactors existed during the Heavy Bombardment epoch, feeding material potentially harbouring viable organisms into near Earth space. During the three stages of planetary reseeding, the encapsulated bacterial population experiences abiotic stressors; specifically, they experience pressure and heat shock twice, in stage 1 and after a recovery phase in stage 2, and again in stage 3. Although many circumstances have played a role in the endurance of life in the early history of the Earth, a particular biological effect could potentially be conferred on a bacterial population in this scenario. Thus, the surviving population not only would experience an increase in the frequency of robust genotypes but also would be expected to have greater stress tolerance than non-stressed organisms of the same species. Hence, in the trampoline effect, the mean robustness of the bacterial population towards these stressors is higher in stage 3 than stage 1. In principle, the time between the impactor and the reimpactor need not be long before this trampoline effect appears. Experiments simulating stage 1 must take this effect into consideration in estimating the survival probabilities of a population of organisms in worlds such as the past Earth. Thus, the stages of planetary reseeding can themselves be considered facilitators of a process that enhances the stress capacity of the collected bacteria and thus their survival capacity. This process may have played a role in the survival of life through violent periods of Earth's history and thus may affect inhabited worlds in general.

**Introduction**

Impactors have hit the Earth ever since its formation via accretion from the solar nebula approximately 4.5 billion years ago. They have continued to be infrequent guests throughout the history of the Earth, although their frequency and size have declined since the Late Heavy Bombardment epoch [Reyes-Ruiza et al., 2012]. The Late Heavy Bombardment occurred approximately 4 billion years ago and is usually considered to have ended 3.8 billion years ago, although evidence indicates that terrestrial impacts did not cease but instead waned gradually until approximately 3.0 Ga ago [Lowe et al., 2014]. Although the Earth remains affected by impactors occasionally, kilometre-scale asteroidal or cometary impactors are now a rare though still potential event estimated to occur at intervals of millions of years [Chapman and Morrison, 1994].

 Although the early part of the Earth's history was characterized by these violent events, life already existed at the time, and evolution therefore was active even then. Thus, although the details remain under debate, and much remains to be elucidated, the first fully autonomous cell with a high certainty existed 3.5 billion years ago [Schopf et al., 2007], in the Archean Eon, which is usually been dated to between 3.8 and 2.5 billion years ago [Coenraads and Koivula, 2007]. The prebiotic processes that led to this cell were probably not a single event; instead, the transition from chemistry to biology was a gradual series of steps of increasing complexity, possibly taking place in a 'great prebiotic spot' on the planet [von Hegner, 2019]. Thus, several lines of evidence point to the emergence of life on Earth between 4.1 to 3.5 billion years ago [Bell et al., 2015], in the Hadean Eon, which usually dates from the end of the Earth's accretion until 3.8 billion years ago [Coenraads and Koivula, 2007].

 With impactors and the presence of life established, life might have been, and still may be, transported between the worlds of the solar system, because such impacts can lead to material being accelerated away from the impact region. Thus, depending on the impactor energy, such ejected material can reach velocities higher than a world's planetary escape velocity, for example, permitting meteorites from Mars to reach Earth and vice versa [Brennecka et al., 2014]. Thus, lithopanspermia is an established hypothesis that proposes a natural exchange in organisms between solar system bodies due to asteroidal or cometary impactors.

The lithopanspermia hypothesis appears to have some justification because billions of rocks were launched during the early violent time of the solar system [Sleep and Zahnle, 1998], and, e.g., Mileikowsky et al. (2000) have estimated that a fraction of a *Bacillus subtilis* spore population ($10^{-6}$) would be capable of surviving radiation in space for approximately 1 million years if shielded by 1 m of meteorite material (assuming the meteorite initially harboured $10^8$ spores/g) and 25 million years if shielded by 2 to 3 m of meteorite material.

Thus, the transport of life from Earth may have reached Venus and even the moons of Jupiter and Saturn. Indeed, life has been suggested to have first appeared on Mars and subsequently been transported to Earth [Davies, 2003]. Lithopanspermia is characterized by three distinct stages in which the transmission of life occurs from a donator world to an acceptor world: (1) planetary ejection, in which organisms must survive ejection from a planet; (2) interplanetary transit, in which organisms must survive the transit between worlds; and (3) planetary entry, in which organisms must survive being deposited on a world [von Hegner, 2019].

In addition to this interesting exchange between solar system bodies, another scenario exists wherein ejected material may instead return to its home planet and 'reseed' life after the world has recovered after the sterilizing effect of a global impactor, in a period that might span decades for Mars or millennia for Earth, thus restarting evolution in a world. This is the so-called 'refugium hypothesis' [Sleep and Zahnle, 1998].

Wells et al. (2003) have calculated that if an impactor ~300 km in diameter with a relative velocity of 30 km s$^{-1}$ were to hit the Earth and eject material into space, a substantial amount of this terrestrial ejecta would in fact return within a time scale of less than 3000–5000 years. Furthermore, if microorganisms had also been launched in this sterilizing global impact, protected within the ejected material, then an initial population of order $10^3$ to $10^5$ organisms kg$^{-1}$ has been estimated to be sufficient for a single organism to endure the refugium and return back to a recovered Earth within this time frame [Wells et al., 2003].

The stages of lithopanspermia are similar in many ways to this scenario, because a viable return to the home planet and viable transfer to a nearby rocky planet have comparable probabilities in terms of the material ejected from a world [Sleep and Zahnle, 1998]. Thus, as for lithopanspermia, three stages of planetary reseeding can be defined: (1) planetary ejection, in which organisms must survive ejection from the planet; (2) planetary proximity refugium, in which organisms must survive the stay in space; and (3) planetary entry, in which organisms must survive being deposited on the world. Thus, early terrestrial life could have survived, e.g., the Late Heavy Bombardment through a temporary shelter within a meteorite in space and subsequently reseeded Earth after an otherwise lethal impact had occurred on the planet.

More local impactors that do not jeopardize the global distribution of life also exist. In fact, in the common scenario, a higher frequency of rocks is in transit due to smaller non-sterilizing impacts, whereas large sterilizing impacts are rare [Sleep and Zahnle, 1998]. Thus, studies have estimated that if an arbitrary impactor size of 1 km diameter were chosen, the current impact rate at that size would be approximately one every 600,000 yrs; this rate would probably have been 100 times higher in the ancient history of the solar system. Hence, a < 6 kyr interval for such impactors would have been the case, with each 1 km impactor capable of ejecting order 1 billion rocks with mean sizes on the order of a decimetre [Miliekowa et. al., 2000; Gladman et al., 2005]. Thus, during the Heavy Bombardment epoch near Earth, space was almost continuously being fed by terrestrial material, owing to the frequent impacts of such smaller impactors, which nonetheless could eject material potentially harbouring viable organisms [Gladman et al., 2005].

The transmission of organisms in stage 2 of lithopanspermia usually requires millions of years, a period that can present significant challenges to the survival of the organisms involved. Indeed, planetary reseeding may thus have a greater probability of organism survival, because an important difference, which will be addressed in this article, is that although organisms must also survive their stay in stage 2, the time for this stage may be relatively briefer. Thus, Gladman et al. (2005) have estimated that approximately 1% of the mass ejected by an impactor with velocity 30 km/s will return to the Earth within 30,000 years. For impactors with ejection velocities between 1 and 2 km/s, a fraction of the ejected mass has been calculated to return after approximately 5000 years. In fact, material ejected with $V_\infty$ = 2 km/s makes up approximately half the mass that returns to Earth, independently of the impactor speed [Gladman et al., 2005]. A study by Reyes-Ruiza et al. (2012) has concluded that particles launched only 1% above Earth's escape velocity, at 11.22 km/s overall, would remain in orbits close to that of the planet, and within 30,000 years, approximately 5% of these particles—7783 out of 163,842 in the simulation—would fall back to Earth, potentially in less than 5,000 years.

Thus, if the conjecture is granted that life can remain viable in all three stages of this planetary reseeding scenario, then this topic relates to not only space science but also biology, thus requiring a framework that integrates the physical dynamics of the stages of planetary reseeding with the dynamics of biological processes, as explored in this article.

**Discussion**

The focus of this work is mainly on the relatively modest impactors/launchers affecting a restricted region of a world, which is probably the dominant scenario in the history of the Earth.

This entire scenario of local bombardments can be symbolically illustrated as a flat surface, i.e., a trampoline. This flat surface is hit by millions of drops of matter spread over time and distance. Each time an impactor hits the trampoline, a drop of matter containing bacteria is sent high above the ground before the drop returns as a reimpactor on the trampoline.

This trampoline or bouncing ball analogy is important, because although many processes and circumstances have played a role in life's endurance and evolution, in the early history of a world like the Earth, a particular biological effect could potentially be conferred on a bacterial population in this scenario.

In the traditional discussion of lithopanspermia and planetary reseeding, life is usually treated as a passive cargo that is enclosed and protected until it arrives in a world and is released again. However, life is not passive but instead reacts to environmental stressors. Thus, evolution occurs, involving a competition for survival between individual phenotypes within any given species, as well as life adapting to changing environments [von Hegner, 2020]. Thus, it can be predicted in evolutionary theory, that robustness—the capacity to endure adverse environmental conditions derived from e.g., cold stress or heat stress in a particular species—can be a trait gradually acquired in response to a tough environment [Lenz et al., 2018].

The three stages of planetary reseeding represent a changed environment for organisms, and the environmental stressors encountered therein thus provoke evolutionary responses. A hypothesis can be formulated in which stress tolerance can also be acquired more rapidly during an invasion process involving members of one species being transported from one region to another over some period of time [Lenz et al., 2018]. In this hypothesis, mortality during transport due to stress will in fact increase the mean tolerance to the same type of stress in the group of invasive organisms. Thus, mortality rates will in fact be lower among survivors of a preceding stress event than among non-stressed organisms belonging to the same species [Gayán et al., 2016; Lenz et al., 2018].

A study by von Hegner (2020) has shown that although the transport of life in lithopanspermia is most likely to occur in a dormant state, a scenario in which life remains active during transmission exists. Here, biological processes play a role, which can endow life with a stress tolerance different from that expected on the basis of analysis of the dynamics through only a planetary science approach. The framework developed therein could also be applied to this similar yet different scenario. Thus, the following assumptions will be made:

(i) The trampoline effect is a prevalent effect among bacteria; i.e., the mean robustness of bacteria towards a stressor is higher at the end of the planetary proximity refugium than at the beginning.
(ii) The bacterial population can remain active in stage 2.

Throughout approximately the first 2 billion years of life's history on this planet, life consisted of single-celled organisms; therefore, only these will be addressed.

**The trampoline effect**

The three stages of planetary reseeding are conjectured to be capable of storing a bacterial (or archaeal) population for a period of time. This storage exposes the encapsulated bacteria to stressful abiotic conditions.

Thus, because organisms can modify their stress tolerance, the question then arises as to what potential processes can occur in the three stages of planetary reseeding.

*Stages 1 and 3*

In planetary reseeding, impactors hit the ground in stage 1 and send up matter containing a bacterial population. Subsequently, in stage 3, a reimpactor returns and delivers the bacterial population to another location in the world.[1] Thus, the bacterial population experiences pressure as well as heat shock.

Initially, one might expect that the few surviving organisms capable of enduring the impactor/launcher in stage 1 can also potentially survive the reimpactor in stage 3, because robustness toward stressors, e.g., heat stress or cold stress in the bacterial population can be an inherent trait that has gradually evolved in response to a harsh environment. Here, life is merely a passive cargo being shipped, stored, and returned by a meteorite [von Hegner, 2020]. However, the situation can be more complex. An interesting biological effect is responsible for the ability of bacterial populations to achieve greater robustness than in this scenario.

When the impactor hits and ejects material from a planet, stage 1 of planetary reseeding results in a shift from a moderate to a momentarily elevated temperature. Hence, during this event, which begins and terminates with a momentarily elevated temperature, all of the individual bacteria within the ejected material experience temperature changes.

Thus, stage 1 involves temperatures that represent a heat shock for the bacterial population. This heat shock can lead to two distinctive consequences of biological significance. First, this heat shock can lead to a decrease in bacterial quantity, meaning that a partial mortality will result among the collected organisms, thereby selecting bacterial genotypes that initially express the greatest tolerance toward the experienced heat shock. Therefore, the quality of the organisms should subsequently increase; that is, the frequency of heat-shock-tolerant genotypes in the bacterial population is expected to increase during stage 2 [von Hegner, 2020]. Second, tolerance to the second stress event, which is of the same type, can potentially be higher than expected solely from the robust genotypes of the bacteria.

The importance of these consequences is illustrated in Figure 1, where bacterial population A is subjected to the three stages of planetary reseeding. Population A endures two pressure- and heat-stress events: one event in stage 1 and—after a reconstitution phase in stage 2, wherein the pressure is alleviated and the heat decreases—another event in stage 3.

In contrast, hypothetical bacterial population B of the same bacterial species endures only one pressure- and heat-stress event, which occurs in stage 3 (due to a virtually non-violent launch through a hypothetical space elevator [Artsutanov, 1960], rather than the violent ejection experienced in stage 1). Populations A and B spend the same amount of time in each meteorite in stage 2 and endure the same conditions in this stage.

The stress experienced by population A during the stage 1 can be expected to cause substantial mortality in the bacterial population. In the discussed scenario, the surviving individuals endure a second stress event of the same type in stage 3. After their survival, the robustness of this population is compared with that of population B, which did not experience a pressure- and heat-stress event prior to stage 3.

Thus, it is predicted that the stress tolerance among population A survivors will be enhanced by the first pressure- and heat-shock event, compared with the stress tolerance of population B. In other words, the robustness of population A toward these stressors is expected to be higher at the end of their refugium in stage 3 than at the beginning of stage 1.

Hence, in stage 1, the impactor/launcher facilitates a natural selection of initially pressure- and heat-tolerant organisms, and this reduction in bacterial quantity increases the quality or average robustness toward the pressure and heat shock occurring in stage 3, due to their previous exposure to these stressors.

Experiments supporting this biological effect have been conducted. Gayán et al. (2016) performed experiments regarding the stress tolerance of heat-stressed *E. coli*. Initially, the organisms that survived were only modestly more sensitive to a subsequent heat stressor and were instead more resistant to a second stressor, a high hydrostatic pressure shock, than unstressed control *E. coli*. However, as the recovery phase proceeded, the initially high hydrostatic pressure resistance of the *E. coli* faded, whereas their heat resistance increased until it surpassed the initial heat resistance of unstressed control *E. coli* [Gayán et al., 2016].

---

[1] In stages 1 and 3 of planetary reseeding, the physical and biological mechanisms are the same as in lithopanspermia; thus, the theory and experiments discussed in the present section generally follow the same presentation as in the section *Stages 1 and 3* in von Hegner (2020).

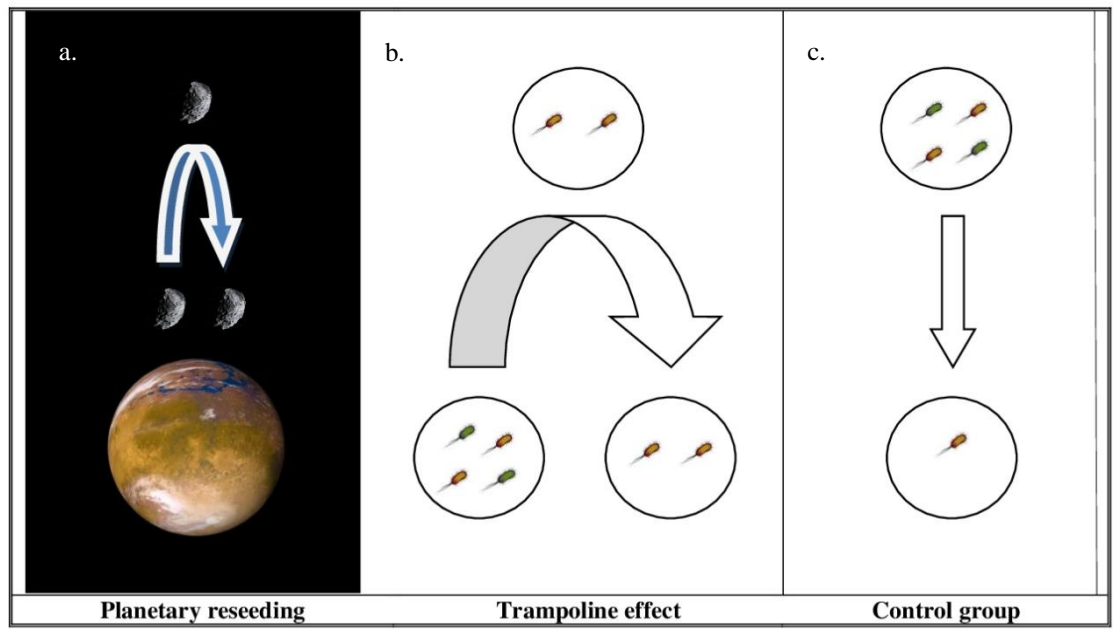

**Figure 1.** (a) The classic scenario in which a bacterial population is transmitted and, after a period of time, returns to the home world again via an asteroid (asteroid is not to scale). (b) Bacterial population A experiences a pressure and heat shock in stage 1 and, after a period in stage 2, returns to the home world with an increased tolerance to a stress event of the same type, which occurs in stage 3. (c) The control bacteria population B experiences only one pressure- and heat-shock event. Credits: artistic image of Ancient Mars by Detlev Van Ravenswaay; bacteria adapted from Mirumur, 2011.

Lenz et al. (2018) conducted a study simulating the heat stress experienced by mussel species during transoceanic transportation. Groups of individual mussels, which are often transported in the ballast water tanks of ships, were subjected to heat stress, which caused 60%–83% mortality in the groups. The surviving organisms subsequently experienced a second heat-stress event in the experiments, and among the organisms, two mussel species displayed an increased tolerance to the subsequent event [Lenz et al., 2018].

Studies of bacterial survival during pressure shocks simulating stage 1 have been conducted by Horneck et al. (2008), who performed recovery experiments on the cyanobacterium *Chroococcidiopsis*. The experiments showed that the number of surviving *Chroococcidiopsis* sp. decreased with increasing shock pressure. Survival was observed at pressures up to 10 GPa (survival rate nearly $10^{-3}$) before approaching the survival below the threshold of detection [Horneck et al., 2008].

These experiments did not test whether the surviving bacteria showed increased robustness toward a second pressure-shock event than a control group experiencing this pressure-shock event only once. Moreover, the number of survivors after stage 1 of these pressure-shock experiments was not necessarily due to an arbitrary survival among the cyanobacteria. Rather, the experiments could select for individual genotypes with a pre-adapted robustness to pressure shock, thus allowing the bacteria expressing the greatest genetically adapted tolerance to survive longer.

However, the trampoline effect is presumably not due to pre-adapted mutants because, as found by Gayán et al. (2016), most heat-stressed *E. coli* survivors did not display significantly increased robustness toward the stressor. Hence, it has been proposed that heat-shock stressors trigger either a translation-independent response or confer an, at present, poorly understood physiological state upon the surviving organisms that mitigates the effects of the second stress event [Gayán et al., 2016].

Thus, bacteria may also exhibit increased pressure-stress robustness to a second stress event of the same type in stage 3 [von Hegner, 2020].

Hence, the bacterial population should not be considered as merely passive cargo stored in the meteorite before reimpact on the planet. Life responds to the environment and thus can affect the outcome of the storage itself as well as successful resettlement on the planet. Therefore, stage 1 can be viewed as a facilitator of a biological process that enhances the ability of the collected organisms to survive stage 3.

*Stage 2*

For the trampoline effect to occur, the meteorite must protect the bacteria in the planetary proximity refugium in stage 2 long enough for them to have a recovery phase and thus actively maintain their most basic functions. The major difference between lithopanspermia and planetary reseeding emerge in this scenario, rather than in the difference between worlds and one world scenarios.

In lithopanspermia, stage 2 acts as a transport and protection of the organisms from a donator world to an acceptor world. In planetary reseeding, the situation is different. Here, the planetary proximity refugium acts as a natural microbial bank, almost analogous to an artificial seed bank, although the organisms are not dormant here.

The possibility of planetary proximity refugium and reseeding appears to be higher than that of lithopanspermia for several reasons. First, the transmission time in stage 2 of lithopanspermia can be long, lasting an average of millions of years, thus reducing the probability of bacterial survival, whereas the time of planetary proximity refugium can be significantly shorter. Second, the transmitted material with organisms has a higher probability of arriving in a potentially habitable world, returning to its home world, than it has of arriving in another potential alien world.

For example, Wells et al. (2003) have calculated that a transmitted meteorite could possibly return to Earth 3000–5000 years later. This calculation was for a global event, but here we restrict the discussion to local non-sterilizing events. During the Heavy Bombardment epoch, smaller impactors would have been dominant. Thus, arbitrary impactors with a size of 1 km diameter, would be estimated to continuously feed near Earth space with each launching on the order of 1 billion viable rocks [Gladman et al., 2005]. Impactors with velocities greater than 2 $V_{esc}$ have been estimated to allow a significant amount of terrestrial material to return in less than 5,000 years [Reyes-Ruiza et al., 2012].

Thus, for planetary reseeding, the time in stage 2 is generally given by:

$$0 < \text{refugium} \leq 5000 \text{ yrs} \qquad (1)$$

where 5000 years is the estimate given by Wells et al. (2003). The authors have argued that in the general hostile conditions of the space environment, e.g., exposure to radiation, biological material would be rendered nonviable after a few thousand years. However, it is important to distinguish between life existing in the refugium period in dormant versus active forms, because the resistance to stressors would differ. Thus, in stage 2, the meteorite must provide an internal protected environment in which the bacterial population can remain active. Many factors affect that possibility. One factor is the meteorite itself, which protects the encapsulated organisms. Another factor is that shorter times spent in space result in greater possibility of staying active. A third factor is the bacterial population itself.

Bacteria are well known to have much faster life cycles than multi-cellular organisms. Thus, the doubling times of bacteria can range from 9.8 minutes to several hours [Eagon, 1962; Gibson et al., 2018]. The recovery or resuscitation time in bacteria after sublethal injury is typically within 2 hours, after which they regain metabolic functions and the ability to divide [Ray, 1986; Kang and Siragusa, 1999]. However, Gayán et al. (2016) have reported that the growth of *E. coli* survivors is evident only approximately 5 h after a heat shock.

In the scenario discussed here, the planetary proximity refugium in stage 2, recovery simply means that the effects of pressure and heat shock on the bacteria decrease, allowing them to resume their functions, thus facilitating increased robustness towards a second stress challenge of the same type in stage 3. The survivors from the impactor do not necessarily need to experience any injury. Thus, if we assume this 5 h observation (in which the effect of pressure shock is not included), then the time for the occurrence of the trampoline effect is given by:

$$5\ h \leq T_{effect} \tag{2}$$

Consequently, in principle, the refugium period need only be within 5 hours for the bacterial population to achieve an enhanced tolerance towards stage 3.

The launch and return times for the impactor/launcher I and reimpactor R in this specific scenario are given by:

$$0 < I < R \leq T_{effect} \tag{3}$$

Thus, in principle, the meteorite need not be sent far away from a world before the trampoline effect appears. In many ways, this situation is opposite from lithopanspermia. Here, the focus is on how far away an impact can launch material, but here there is only 5 h between the impactor and the reimpactor. The material may not even need to be ejected with an escape velocity greater than that of the planet.

This short period of time is not an issue for another reason. First, this is not a situation in which life can return only after a world has recovered from a global impactor. Because while a local impactor/launcher has sterilized the affected area, the local reimpactor probably would not hit the same spot as the impactor/launcher. Instead, it would hit another location on the planet where life or the possibility for life can exist. Of course, when the reimpactor hits, it could also sterilize the area. However, the surrounding area could still remain fertile. Second, a main point is that the returning organisms have a greater tolerance to a stress event of the same type in stage 3 than they had when they were ejected in stage 1. Thus, they can actually demonstrate higher tolerance towards pressure and heat shock than they did before.

Thus, the first and second stages of planetary reseeding can in themselves be considered facilitators of a process that enhances the stress capacity of the collected bacteria and thus their survival probability.

*The stages of planetary reseeding: biological processes*

The focus here has been on local events rather than global sterilizing events, although the trampoline effect may also affect planet-wide impactors/launchers. However, such local events have been among the dominant scenarios, which are produced with greater frequency, although global events may have occurred during the early history of the Earth.

Such repetitive local impacts may also have implications beyond those of the global impactors/launchers. Thus, repetitive local impacts may have played a greater role beyond only reseeding life after a single large global impact, because repeated inputs of organisms with different tolerances may have significance in evolution. Darwinian evolution is, after all, based on competition for survival among individual organisms within any given species. Thus, optimal trait values change continuously, and natural selection results in gradual evolution of fitter organisms, hence removing suboptimal forms that hitherto were well adapted to a given environment [Simons, 2011].

Repeated local reimpactors delivering inputs of organisms with not only enhanced stress tolerance but also different genotypes would therefore give organisms a better chance of surviving in the violent period of a world than if a uniform population of organisms arrived at once after a global impactor. Thus, if the time between the local impactor and local reimpactor is measured in years, or the local impactor and local reimpactor arrive at widely different places in a world, then genetic diversity is affected: the arriving bacteria are placed among bacteria that remained on the planet and are thus different from them, and a competition between them therefore occurs.

However, the organisms are still assumed to arrive in an environment that is sufficiently similar to the environment from which they came, so that they can survive the initial encounter there. If the environment is too different, then the chances of survival diminish, even if the environment is habitable [von Hegner, 2019].

Of course, the stress tolerance or robustness achieved is specific to only the stress experienced by the bacterial population through these stages. Tolerance would not necessarily be conferred towards all other types of environmental stressors present in a world, such as on the early Earth, encountered by the bacteria upon their return. However, if the meteorite protects the bacteria well enough for them to be active, and they have sufficient time, then they can also develop stress hardening [Lou and Yousef, 1996] and cross-tolerance [Johnson, 2002], which could confer robustness towards many different stressors.

However, organisms living on the planet may also develop stress hardening and cross-tolerance to the environmental stressors that they experience, and the arriving bacteria may not necessarily have a competitive advantage. Therefore, the arriving bacteria may not necessarily gain benefits that the organisms that remained on the planet could not also achieve.

If the bacteria remain active after their recovery, nutrients are sufficient, and the bacteria are protected in the meteorite for 3000 to 5000 years before their return, then another scenario would emerge. The population is not increased in the scenario discussed to date, but if the bacterial population were able to remain active in the meteorite for, e.g., 3000 years, then their numbers would increase again, and evolution would thus be in effect. Thus, the quantity of the bacterial population could initially decrease through stage 1 of planetary reseeding but subsequently increase again through stage 2, in effect distributing quality along the way; therefore, the quality of the population would increase. In addition, the bacterial population could potentially achieve even greater survival capacity before entering stage 3, because natural selection among the organisms in the meteorite environment could result in the evolution of new traits not previously possessed.

Furthermore, because life, or more precisely, survival, is the primary topic of interest, rather than the physical dynamics of impactors/launchers and reimpactors, a meteorite that successfully ejects, stores, and returns life, can itself be considered a physical invariant.

Thus, the stages in planetary reseeding can themselves be considered facilitators of a process that enhances the biological survival capacity of the collected organisms.

**Conclusion**

The overall situation discussed in this article is that ejected material can return to Earth again after 3000 to 5000 years. This is a short period from an astronomical point of view but a long period from a biological point of view. Conversely, there is the situation that surviving bacteria can recover from stress after 2 to 5 hours. This duration is scarcely noticeable from an astronomical point of view, but from a biological point of view, much can occur during that time span.

In the overall situation discussed, stress-induced mortality in stage 1 can potentially increase quality or population robustness towards a second stress challenge of the same type in stage 3 of planetary reseeding. This response can be illustrated as a trampoline effect. The bacteria on the trampoline can better endure the reimpactor than they could when they were first ejected by the impactor. Thus, although the survival probability of bacteria can be estimated through, e.g., pressure shock experiments simulating stage 1, the next stages will give a probability of survival higher than would otherwise be expected from this type of experiment; this aspect must be taken into consideration when attempting to estimate the survival probabilities of a population of organisms in worlds such as the past Earth.

Of course, this trampoline effect does not represent an adaptive *perpetuum mobile*. A repeated series of transmission and return of the same bacterial population could not be expected to lead to a population of 'super' organisms capable of withstanding any and all degrees of environmental stress. Evolutionary adaptation would be necessary to truly evolve the organisms further. However, from the few experiments performed to date, how far such a repeated series can proceed is not yet known but would certainly be interesting to know.

Whether the impact of drops by local impactors/launchers and reimpactors collectively has had a bearing on driving the survival of life on a flat surface is an open question. This specific adaptive scenario remains among other scenarios. What is known is that besides the infrequent bombardment itself, the Hadean and Archean environment on Earth was in many ways a more harsh and stressful environment than today's Earth. Upon launching, the bacterial population would not only ensure the continuation of life after a potential sterilizing event but, upon return, the bacterial population would have greater tolerance to pressure and heat stress than they had when ejected, thereby simultaneously increasing their survival capacity. Thus, this process may have played a role in the survival of life through these violent periods of Earth's history. The importance of this effect in life's early history is unknown, but its existence appears to be clear. This effect may thus influence inhabited worlds in general.